\documentclass[aps,prb]{revtex4}
\usepackage{graphicx}
\usepackage{soul}

\begin{document}
\pagestyle{empty}
\title{Influence   of electric current on the  Casimir forces between graphene sheets }

\author{A.I.Volokitin$^{1,2}$\footnote{Corresponding author.
\textit{E-mail address}:alevolokitin@yandex.ru}    and B.N.J.Persson $^1$}
 \affiliation{$^1$Peter Gr\"unberg Institut,
Forschungszentrum J\"ulich, D-52425, Germany} \affiliation{
$^2$Samara State Technical University, 443100 Samara, Russia}

\begin{abstract}
We investigate the dependence of the thermal Casimir force and the Casimir friction force between two graphene sheets on the drift velocity of the electrons in one graphene sheet. We show that the drift motion  produces a measurable change of the thermal Casimir force due to the Doppler effect. The thermal Casimir force as well as the Casimir friction are strongly enhanced in the case of resonant photon tunneling when the energy of the emitted photon coincides with the energy of electron-hole pair excitations.   In the case of resonant photon tunneling, even for  temperatures above room temperature the Casimir friction is dominated by quantum friction due to quantum fluctuations. Quantum friction can be detected in frictional drag experiment between graphene sheets for high electric field.
\end{abstract}

\maketitle


\vskip 5mm

In the late 1940s Hendrik Casimir predicted \cite{Casimir1948} that two macroscopic non-magnetic bodies with
no net electric charge (or charge moments) can experience an attractive force much stronger than gravity. The existence of this force is one of the few direct macroscopic manifestations of quantum mechanics; others are superfluidity, superconductivity,  and the black
body radiation spectrum.

 Casimir based his prediction on a simplified model involving two parallel perfectly conducting plates separated by  vacuum.
A  theory of  the van der Waals and Casimir forces between  parallel material plates in
thermal equilibrium and separated by a vacuum gap was developed by Lifshitz (1955)\cite{Lifshitz1955}.   Lifshitz's theory describes
dispersion forces between dissipative media as a physical phenomenon caused by the fluctuating electromagnetic
field that is always present in both the interior and the  exterior of any medium. Outside the medium this field exists partly in the
form of the radiative propagating waves, and partly in the form of nonradiative evanescent waves whose amplitudes decay exponentially with
the distance away from the medium. To calculate the fluctuating electromagnetic field Lifshitz used Rytov's theory \cite{Rytov1953,Rytov1967,Rytov1989}. Rytov's theory is based
on the introduction into the Maxwell equation of a ``random'' field (just as, for example, one introduces a ``random'' force in the theory of Brownian motion).  Both quantum and thermal fluctuations give contributions to the total Casimir force. A general theory of Casimir and van der Waals forces was developed in Ref. \cite{Dzyaloshinskii1961} using quantum field theory. This theory confirmed the results of Lifshitz's theory. Quantum fluctuations dominate at small separation
 ($d< \lambda_T=c\hbar/k_BT$) and thermal fluctuations  dominate at large separation ($d> \lambda_T$). Casimir forces due to quantum fluctuations
  have  been studied experimentally for a long time \cite{Lamoreaux2007,Milloni1994}. However the Casimir forces due to thermal fluctuations were measured only recently, and the results confirmed the prediction of the Lifshitz theory \cite{NaturePhysics2011}. At present the interest in Casimir forces is increasing
 because they dominate the interaction between nanostructures and are responsible for  the adhesion between moving parts in small devices such as
 micro- and nano-electromechanical systems \cite{Serry1998,Buks2001}. Due to this practical interest and the fast progress in  force detection
 techniques, experimental \cite{NaturePhysics2011,Klimchinskaya2009,Munday2009,Sushkov2011,Bao2010} and theoretical \cite{Johnson2011,Zhao2009} investigations  of Casimir forces have experienced an extraordinary ``renaissance'' in the past few years. At present  a great deal of attention is devoted to the study of the Casimir forces in  graphene  systems \cite{Bordag2006,Bordag2009,Drosdoff2010,Drosdoff2011,Fialkovsky2011,Sernelius2011,Svetovoy2011,Sarabadani2011,Gomez2009}.

 Graphene, isolated monolayer of carbon, which was  obtained very recently, consists of carbon atoms densely packed into a two-dimensional honeycomb crystal lattice. The unique electronic and mechanical properties of graphene are actively studied both theoretically, and experimentally because of their importance for fundamental physics, and also possible technological applications {\cite{Geim2004, Geim2005,Geim2007,Geim2009}. In particular, the valence band and conduction band in graphene touch each other at one point named  the Dirac point. Near this point the energy spectrum for electrons and holes has a linear dispersion. Due to this linear (or ``conical") dispersion relation  electrons and holes near this point behave like relativistic particles described by the Dirac equation for massless fermions.
Because of the unusual electronic properties of graphene, the Casimir forces in graphene also have unusual properties. Contribution to the Casimir force due to thermal fluctuations for normal materials dominates for $d>\gamma_T=\hbar c/k_BT$ but for two graphene sheets thermal contribution dominates for much shorter distances \cite{Gomez2009} $d>\xi_T=\hbar v_F/k_BT$ , where $v_F\sim 10^6$ m/s is the Fermi velocity in graphene. At room temperature the parameters   $\xi_T$ and $\gamma_T$ are 25 nm and 7.6 $\mu$m, respectively. This property  makes it possible to measure the thermal Casimir force using an atomic force microscope or other force measuring techniques. Tailoring the thermal Casimir force using Fermi level tuning by gate voltage was discussed in Ref. \cite{Svetovoy2011}.

Alternative method of tailoring  the thermal Casimir force consists in driving  an electric current in a graphene sheet. It was shown by Pendry} \cite{Pendry1997} that the reflection amplitudes from moving metal surface
are modified due to the Doppler effect. The same modification of reflection amplitudes can be obtained if instead of motion of  metal plate, a drift motion of charge carriers is induced in it by applied voltage \cite{Shapiro2010}.
 Due to the high mobility of carriers in graphene, in a high electric field electrons (or holes)  can move with very high velocities (up to $10^6$ m/s). The drift motion  of charge carries in graphene will result in a modification of dielectric properties (and the Casimir force) of graphene  due to  the Doppler effect \cite{Pendry1997}. If in one of   two parallel graphene sheets an electric current is induced, then the electromagnetic waves, radiated by the graphene sheet  without  an electric current, will experience a frequency  Doppler  shift  in   the reference frame moving with the drift velocity $v$ of electrons in the other graphene sheet: $\omega^{\prime}=\omega+q_xv$, where $q_x$ is the parallel to the surface component of momentum transfer. The same is true for the waves emitted by the other graphene sheet. Due to the frequency dependence of the reflection amplitudes the electromagnetic waves will reflect differently in comparison to the case when there is no drift motion of electrons, and this will give rise to the change of the Casimir force.

In this Letter, we investigate the dependence of the thermal Casimir force between graphene sheets on
the drift velocity $v$ of charge carriers in one of the graphene sheet.
 Let us consider two  graphene sheets  separated by vacuum gap with thickness $d\ll \lambda_T=c\hbar/k_BT$.
Assume that the free charge carriers in one graphene sheet move  with drift
velocity $v\ll c$  along the $x$-axis ($c$ is the light velocity) relative to the other graphene sheet. Because a drift motion of the free charge carriers produces a similar modification of the reflection amplitudes as in the case of moving graphene sheet, the theory of the Casimir forces between moving  bodies \cite{Volokitin2008b} can be used to calculate the Casimir forces between sheets (both of which are at the rest) in presence of the drift motion of the free charge carriers in one graphene sheet.  The force which acts  on the surface of the sheet can be calculated from the Maxwell stress tensor
$\sigma_{ij}$, evaluated at the surface of the sheet at $z=0$:
\[
\sigma _{ij} =\frac 1{4\pi }\int_0^\infty d\omega \int \frac{d^2q}{(2\pi)^2}
\Big[ <E_iE_j^*> + <E_i^*E_j> + <B_iB_j^*> + <B_i^*B_j>
\]
\begin{equation}
 - \delta_{ij}(<\mathbf{E\cdot E^*}> + <\mathbf{B\cdot B^*}>)\Big] _{z=0}   \label{stress}
\end{equation}
where  $<...>$ denotes statistical average over the random the electric $\mathbf{E}$ and magnetic induction $\mathbf{B}$ field.
According to Ref.
\cite{Volokitin2008b} the Casimir force $F_z=\sigma _{zz}$ between moving media  is determined by

\begin{equation}
F_z=F_{zT}+F_{z0},
\end{equation}
where the temperature dependent term $F_{zT}$ and the zero-temperature contribution $F_{z0}$ are given by
\[
F_{zT} =\frac \hbar {\pi ^3}\int_{0 }^\infty dq_y\int_0^\infty
dq_xqe^{-2qd}\Bigg \{ \int_0^\infty d\omega \Bigg(
\frac{\mathrm{Im}R_{1}(\omega)\mathrm{Re}R_{2}(\omega^+)n_1(\omega) + \mathrm{Re}R_{1}(\omega)\mathrm{Im}R_{2}(\omega^+) n_2(\omega^+)}{\mid
1-e^{-2 q d}R_{1}(\omega)R_{2}(\omega^+)\mid ^2}
\]
\begin{equation}
 + (1 \leftrightarrow 2)\Bigg )+ \int_0^{q_xv}d\omega \Bigg(\frac{\mathrm{Re}R_{1}(\omega^-)\mathrm{Im}
R_{2}(\omega)n_2(\omega)} {\mid 1-e^{-2qd}R_{1}(\omega^-)R_{2}(\omega)\mid
^2}  + (1 \leftrightarrow 2)\Bigg )\Bigg \}, \label{CasimirT}
\end{equation}

\[
F_{z0} =\frac \hbar {2\pi ^3}\int_{0 }^\infty dq_y\int_0^\infty
dq_x\Bigg \{ \mathrm{Re}\int_0^\infty d\omega se^{-2sd}\Bigg(
\frac{R_{1}(i\omega)R_{2}(i\omega+q_xv)  }{
1-e^{-2 s d}R_{1}(i\omega)R_{2}(i\omega+q_xv)}
\]
\begin{equation}
 + (1 \leftrightarrow 2)\Bigg )+\int_0^{q_xv}d\omega qe^{-2qd}\Bigg(\frac{\mathrm{Im}R_{1}(\omega)\mathrm{Re}
R_{2}(\omega^-) } {\mid 1-e^{-2qd}R_{1}(i\omega)R_{2}(\omega^-)\mid ^2} + (1 \leftrightarrow 2)\Bigg )
  \Bigg \}, \label{CasimirZero}
\end{equation}
where   $n_i(\omega )=[\exp (\hbar \omega /k_BT_i)-1]^{-1}$
($i=1,2$), $q=\sqrt{q_x^2 + q_y^2}$, $s=\sqrt{(\omega/c)^2+q^2}$, $T_i$ is the temperature of $i$-th graphene sheet,
 $R_{i}$  is the reflection amplitude for
surface $i$ for $p$ -polarized electromagnetic waves, and
$\omega^{\pm}=\omega \pm q_xv$.  The symbol $(1 \leftrightarrow 2)$ denotes the terms that are obtained from
the preceding terms by permutation of $1$ and $2$. In the first term in Eq. (\ref{CasimirZero})  the integration along the real axis was transformed into integration along the imaginary axis.

The reflection amplitude for a 2D-system    is determined by \cite{Volokitin2001b}
\begin{equation}
R_i=\frac{\epsilon _i-1}{\epsilon _i+1}, \,\,\, \epsilon_i=\frac{4\pi p \sigma_i}{\omega \varepsilon}+1,
 \label{refcoef}
\end{equation}
where $p=\sqrt{(\omega/c)^2-q^2}$, $\sigma_i$ is the longitudinal conductivity of the sheet which can written in the form $\sigma_i=-i\omega e^2\Pi_i(\omega,q)/q^2$ where  $\Pi_i$ is the 2D polarizability.  The dielectric function of the sheet is determined by $\varepsilon_i(\omega,q)=1+v_q\Pi_i(\omega,q)$, $v_q=2\pi e^2/q$ is the 2D Coulomb interaction. In term of $\varepsilon_i$ the reflection amplitude can be written as
\begin{equation}
R_i=\frac{p(\varepsilon _i-1)}{p(\varepsilon _i-1)+iq}
 \label{refcoef1}
\end{equation}
In the integration on the real axis $p\approx iq$ for $d<\lambda_T$. Thus, in this case
\begin{equation}
R_i\approx\frac{\varepsilon_i-1}{\varepsilon _i},
 \label{refcoef2}
\end{equation}
On the imaginary axis $p=is$. In the finite lifetime generalization according to the Mermin approximation \cite{Mermin1970} the dielectric function is determined by
\begin{equation}
\varepsilon(\omega,q) \approx 1 + \frac{(\omega + i\gamma)(\varepsilon_0(\omega + i\gamma,q)-1)}{\omega +i\gamma (\varepsilon_0(\omega + i\gamma,q)-1)/
(\varepsilon_0(0,q)-1)},
\end{equation}
where $\varepsilon_0(\omega, q)$ is the RPA dielectric function and $\gamma$ is the damping parameter.
In the study below we used the dielectric function of graphene, which was
calculated recently within the random-phase approximation (RPA)
\cite{Wunsch2006,Hwang2007}. The small (and constant) value of
the graphene Wigner-Seitz radius $r_s$ indicates that it is a weakly
interacting system for all carries densities, making the RPA an
excellent approximation for graphene (RPA is asymptotically
exact in the $r_s\ll1$ limit).  The dielectric function
is an analytical function in the upper half-space
of the complex $\omega$-plane:
\begin{equation}
\varepsilon_0(\omega,q)=1+\frac{4k_Fe^2}{\hbar
v_Fq}-\frac{e^2q}{2\hbar \sqrt{\omega^2-v_F^2q^2}}\Bigg \{G\Bigg
(\frac{\omega+2v_Fk_F}{v_Fq}\Bigg )- G\Bigg
(\frac{\omega-2v_Fk_F}{v_Fq}\Bigg )-i\pi \Bigg \},
\end{equation}
where
\begin{equation}
G(x)=x\sqrt{x^2-1} - \ln(x+\sqrt{x^2-1}),
\end{equation}
where the Fermi wave vector $k_F=(\pi n)^{1/2}$, $n$ is the
concentration of charge carriers, the Fermi energy
$\epsilon_F=\hbar v_Fk_F$,  $v_F\approx 10^6$ m/s is the Fermi velocity. The damping parameter $\gamma$ is due to scattering against  impurities and acoustic phonons  in graphene sheet,
and can be expressed through the low field mobility $\mu $: $\gamma = ev_F/(\hbar k_F\mu$). Scattering of the graphene carries by the acoustic phonons
 of graphene places an intrinsic limits on the low-field room temperature ($T_0=300$ K) mobility, given by $\mu_0=$20 m$^2$/Vs at the graphene carriers density $10^{16}$ m$^{-2}$ (see Ref. \cite{Chen2008}), which gives $\gamma=8\cdot 10^{11}$ s$^{-1}$. At other temperatures the mobility can be obtained using the relation $\mu =
 \mu_0T_0/T$.

In addition to the intrinsic friction due to scattering against  impurities and phonons, on the  electrons moving in the graphene sheet acts the extrinsic friction  due to the interaction with electrons in the nearby graphene sheet. According to the theory of the Casimir friction \cite{Volokitin2008b}, the friction force $F_x=\sigma _{xz}=F_{xT}+F_{x0}$, where at $d\ll\lambda_T$ and $v\ll c$ the contributions from thermal ($F_{xT}$) and quantum ($F_{x0}$) fluctuations are given by \cite{Pendry1997,Volokitin1999,VolokitinRMP2007,VolokitinPRL2011}

\[
F_{xT} =\frac \hbar {\pi ^3}\int_{0 }^\infty dq_y\int_0^\infty
dq_xq_xe^{-2qd}\Bigg \{ \int_0^\infty d\omega \Bigg(
\frac{\mathrm{Im}R_{1}(\omega)\mathrm{Im}R_{2}(\omega^+) }{\mid
1-e^{-2 q d}R_{1}(\omega)R_{2}(\omega^+)\mid ^2}\times
 [n_1(\omega )-n_2(\omega^+)]+(1\leftrightarrow 2)\Bigg )
\]
\begin{equation}
 -\int_0^{q_xv}d\omega \Bigg(\frac{\mathrm{Im}R_{1}(\omega)\mathrm{Im}
R_{2}(\omega^-)} {\mid 1-e^{-2qd}R_{1}(\omega)R_{2}(\omega^-)\mid
^2} n_1(\omega) +(1\leftrightarrow 2)\Bigg )\Bigg \}, \label{FrictionT}
\end{equation}
\begin{equation}
F_{x0} =-\frac \hbar {2\pi ^3}\int_{0 }^\infty dq_y\int_0^\infty
dq_xq_xe^{-2qd}\int_0^{q_xv}d\omega \Bigg(\frac{\mathrm{Im}R_{1}(\omega)\mathrm{Im}
R_{2}(\omega^-)} {\mid 1-e^{-2qd}R_{1}(\omega)R_{2}(\omega^-)\mid
^2}  +(1\leftrightarrow 2)\Bigg ).\label{Friction0}
\end{equation}

 Eqs. (\ref{FrictionT}) and (\ref{Friction0}) were initially obtained for 3D-systems in Ref. \cite{Pendry1997} at $T=0$ and in \cite{Volokitin1999}) for finite temperatures. However, in Ref. (\cite{Volokitin2001b}) it was shown that the same formulas are valid for 2D-systems.
For $v<dk_BT/\hbar $ (at $d=1$ nm and $T=300$ K for $v<4\cdot 10^4$ m/s)
the main contribution to the friction (Eq. (\ref{FrictionT}))
depends linearly on the sliding velocity $v$ so that   the
friction force $F_{xT} = \Gamma v$ where at $T_1=T_2=T$ the friction coefficient $\Gamma$  is given by

\begin{equation}
\Gamma=\frac{ \hbar^2} {8\pi ^2k_BT}\int_0^\infty \frac{d\omega}{\mathrm{sinh^2}\left(\frac{\hbar \omega}{2k_BT}\right)}
 \int_{0}^\infty
dq\,q^3e^{-2qd} \frac {\mathrm{Im}R_{1}(\omega)\mathrm{Im}R_{2}(\omega)}{\left|
1-e^{-2qd}R_{1}(\omega)R_{2}(\omega)\right| ^2}.  \label{parallel6}
\end{equation}

\begin{figure}
\includegraphics[width=0.90\textwidth]{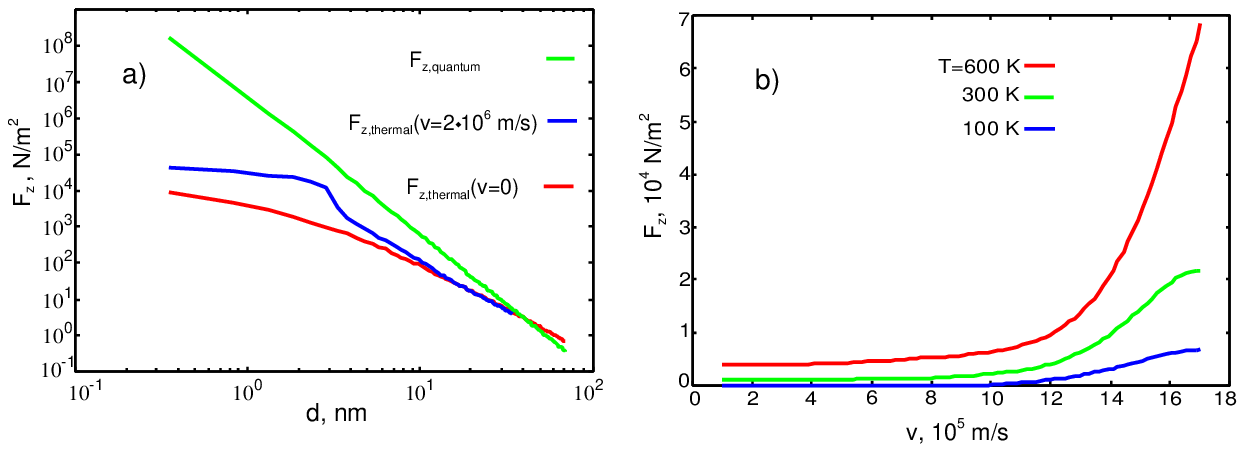}
\caption{\label{Fig1} The Casimir forces between two graphene sheets with carrier concentration $n=10^{16}$m$^{-2}$.   (a) The dependence of the Casimir force on the separation $d$ between the sheets. The thermal and quantum contributions to the total Casimir force are shown separately. The thermal contribution is shown for $T=600$ K and for  the drift velocities $v=0$ and $v=2\cdot 10^6$ m/s.  (b) The dependence of the thermal Casimir force   on the drift velocity of electrons $v$ in one of the graphene sheet at $d=1$ nm.
}
\end{figure}

Due to the presence of an exponential factor in the expression (\ref{CasimirT}) for the thermal  contribution to the Casimir force, the integration over frequency is effectively limited to $\omega < \omega_T=k_BT/\hbar$. Thus for $q_xv\sim v/d > \omega_T$ (at room temperature and for $d=1$ nm this condition corresponds to the velocities $v>10^5$ m/s) the integrand will be modified in the whole range of integration, which will give rise to the significant change of the thermal Casimir force. This change will be especially large in the case of resonant photon tunneling when the integrand has sharp resonances.  The integrand in the expression for the zero-temperature contribution to the Casimir force does not contain any sharp cut-off in the frequency integration. Thus the range of integration will be more wide and the change of the zero-temperature contribution will be significant only for much higher velocities then for the thermal contribution.

Fig. \ref{Fig1}a shows the dependence of the Casimir force between two graphene sheets on the separation $d$ between the sheets.  The thermal and quantum contributions are shown separately. The thermal contribution was calculated for $T=600$ K and for the drift velocities $v=0$ and $v=2\cdot 10^6$ m/s. The thermal contribution becomes larger then the quantum contribution for $d>50$ nm. For $d<5$ nm the thermal contribution calculated for $v=2\cdot 10^6$ m/s is significantly larger then the thermal contribution calculated at $v=0$. For example, at $d\approx 3$ nm the drift motion of the electrons gives rise to the increase of the thermal Casimir force by one order of magnitude, and in this case the thermal contribution is only one order of magnitude smaller then the quantum contribution, and can be measured experimentally. Figure \ref{Fig1}b  shows the dependence of the thermal Casimir force $F_{zT}$  on the drift velocity of the electrons in the graphene sheet at $d=1$ nm. Note the significant change of the thermal Casimir force  for $ v/d > \omega_T$ (at room temperature and for $d=1$ nm this condition corresponds to the velocities $v>10^5$ m/s). This change is connected with resonant photon tunneling. In this case  the photon emitted by the moving electron system (with energy $\omega_{ph}(q)=q_xv-\omega_{eh}(q)$, where $\omega_{eh}(q)$ is the energy of the electron-hole pair excitation with momentum $\mathbf{q}$, will create excitation with energy $\omega_{eh}(q)$ in other graphene sheet.  In the case of graphene  the energy of the electron-hole pair excitation  $\omega_{eh}(q)\approx v_Fq$, where $v_F$ is the Fermi velocity. Resonance occurs when $q_xv\approx 2v_Fq$, which corresponds to $v>2v_F\approx 2\cdot 10^6$ m/s, in accordance with the  numerical calculations.

\begin{figure}
\includegraphics[width=0.70\textwidth]{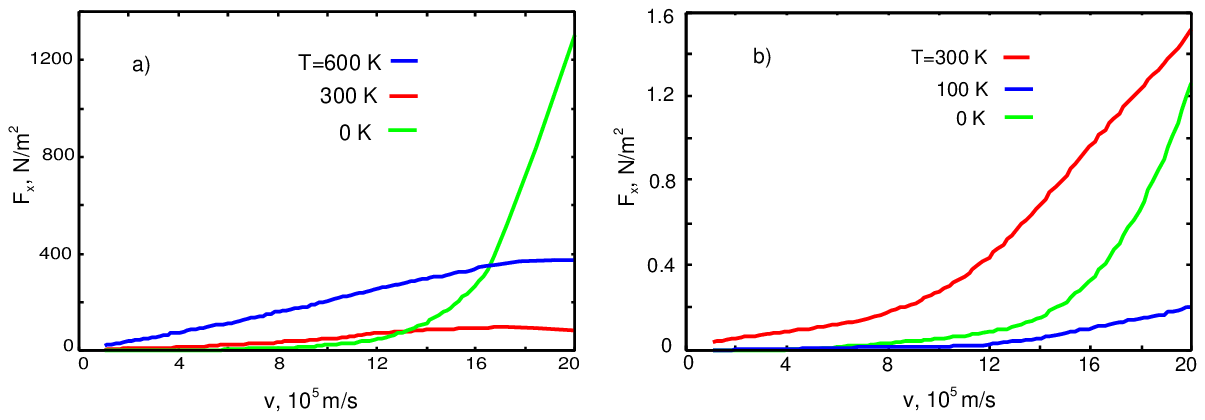}
\caption{\label{Fig2} The Casimir friction force between two graphene sheets
at the carrier concentration $n=10^{12}$ cm$^{-2}$.  The finite temperature curves show only the thermal contributions to the friction. (a) Dependence of
friction force between graphene sheets on the drift velocity of charge
carriers in one graphene sheet at the layer separation $d=1$ nm.
(b) The same as in (a) but at $d=10$ nm.}
\end{figure}

The Casimir friction force between two graphene sheets can be measured in frictional drag experiment. Such experiment was proposed theoretically some years ago  \cite{Pogrebinskii1977,Price1983} and performed experimentally for 2D-quantum wells \cite{Gramila1991,Sivan1992}. In these
experiments a current  is driven through layer \textbf{1}.  Due to the proximity of the layers, the interlayer interactions will induce a current in layer \textbf{2} due to a friction stress  acting on the electrons in the layer \textbf{2} from layer \textbf{1}. If the layer \textbf{2} is an open circuit, an electric field $E_1$ will develop in the layer whose influence cancels the frictional stress $\sigma$ between the layers. In one experiment \cite{Gramila1991} the drift velocity $v\sim 10^2$ m/s. According to the theory of the Casimir friction \cite{Volokitin1999,Volokitin2001b}, at such velocities the thermal fluctuation give the dominant contribution to the friction, and the theoretical predictions are in an agreement with experiment.

Frictional drag between graphene sheets was measured recently in Refs. \cite{Kim2011,Geim2012}. This study has fueled the recent theoretical investigations of frictional drag between graphene sheets  mediated by a fluctuating Coulomb field \cite{Tse2007,Katsnelson2011,Peres2011,Hwang2011,Narozhny2012,Katsnelson2012,Amorin2012}. All these investigations assumes that the friction depends  linear to the current density (or drift velocity $v$ of the charge carries).  Thus only  the thermal contribution to the frictional drag was included. In the linear approximation the electric field induced by the frictional drag depends linearly on the current density $J=nev$ (or drift velocity $v$ of the charge carries), $E=\rho_DJ=F_{xT}/ne=\Gamma J/(ne)^2$, where $\Gamma$ is the friction coefficient, $\rho_D=\Gamma/(ne)^2$ is the drag resistivity. For $\omega < v_Fq$ and $q<2k_F$ the dielectric function of graphene has the following form \cite{Hwang2007}
\begin{equation}
\varepsilon_0(\omega, q)\approx 1+\frac {4e^2k_F}{\hbar v_Fq}\left(1+i\frac {\omega}{v_Fq}\right), \label{lowfrequencydielectric}
\end{equation}
and the reflection amplitude
\begin{equation}
R_0(\omega, q)=\frac{\varepsilon_0(\omega, q)-1}{\varepsilon_0(\omega, q)}\approx 1+i\frac {\hbar\omega}{4e^2k_F}, \label{lowfrequencyreflection}
\end{equation}
and   Eqs.(\ref{lowfrequencyreflection}),
(\ref{parallel6}) give the known result \cite{Tse2007}
\begin{equation}
\rho_D=\frac{\Gamma}{(ne)^2}=\frac{h}{e^2}\frac{\pi\zeta(3)}{32}\left(\frac{k_BT}{\epsilon_F}\right)^2\frac{1}{(k_Fd)^2}\frac{1}{(k_{TF}d)^2},
\label{dragresistivity}
\end{equation}
where $k_{TF}=4e^2k_F/{\hbar v_F}$ is the Thomas-Fermi  screening wave vector.
The frictional drag force is much higher for high drift velocities ($\sim 10^6$ m/s), where it depends nonlinearly  on the drift velocity, and is dominated by the quantum friction, existence of which was  recently  hot debated
\cite { Philbin2009,Pendry2010a,Leonhardt2010,Pendry2010b,Volokitin2011,Philbin2011}.  For $v<v_F$ Eqs.(\ref{lowfrequencyreflection}),
(\ref{Friction0}) give the following  result for quantum friction
\begin{equation}
F_{x0}=\frac{\hbar v}{d^4}\frac{15\zeta(5)}{128\pi^2}\left(\frac{v}{v_F}\right)^2\frac{1}{(k_{TF}d)^2}.
\label{2Friction0}
\end{equation}

In linear approximation  $E=5\times 10^{-4}v$ (SI-units) for $T=300$ K and $d=10$
nm. For a graphene sheet of length $1\ {\rm \mu m}$, and with $v=100$ m/s this electric field will induce the voltage $V=10$ nV.From Eqs.(\ref{dragresistivity}) and (\ref{2Friction0}) the ratio of quantum and thermal friction $F_{x0}/F_{xT}=F_{x0}/(ne)^2\rho_Dv\approx (15/8\pi^2)(v/v_T)^2$, where $v_T=\omega_Td$. Thus, for $v>v_T$ the friction is dominated by quantum friction (at $d=1$ nm and room temperature: $v_T\approx4\cdot10^4$m/s),

Figures \ref{Fig2}a  and \ref{Fig2}b show that much larger electric fields can be induced  at $d=1$ nm (a) and
$d=10$ nm (b) at large velocities. In these figures the contributions to friction from thermal and quantum fluctuations are shown separately, where
the friction force is related to the electric field: $F_x=neE$.  For $v < 10^5$ m/s the frictional drag effect   for the graphene sheets  strongly
depends on temperature, i.e. it is determined mainly by the thermal fluctuations. However, for $v > 10^6$m/s it will be dominated by quantum fluctuations.
Strong enhancement of friction occurs in the case of resonant photon tunneling. As  discussed above,  resonant photon tunneling occurs for $v>2v_F\approx2\cdot 10^6$ m/s. For such velocities
and $d=1$ nm quantum friction dominates over the thermal contribution even at room temperature (see Fig.\ref{Fig1}a). For $d=10$ nm quantum friction dominates at low temperatures (see Fig.\ref{Fig2}b).
 As  discussed in Ref. \cite{VolokitinPRL2011} quantum friction can be also detected by measuring the transport properties of non-suspended graphene on  a SiO$_2$ substrate.

 \textit{Concluding remarks}.--We have calculated the dependence of the thermal Casimir force between two graphene sheets on the drift velocity of the charge carriers  in one graphene sheet. We have found that the drift motion of the charge carriers in graphene produces changes in the thermal Casimir force which can be measured experimentally. The thermal Casimir force, as well as the Casimir friction force, are strongly enhanced in the case of resonant phonon tunneling. For resonant photon tunneling and small $d$, even for  temperatures above room temperature, the Casimir friction is dominated by quantum friction due to quantum fluctuations. Quantum friction can be detected in friction drag experiment between graphene sheets. Another way to detect quantum friction consists in measuring of the transport properties of nonsuspended graphene on an SiO$_2$ substrate in the high electric field.

 \vskip 0.5cm
\textbf{Acknowledgment}

A.I.V. acknowledges financial support from  Russian Foundation for Basic Research (Grant N 12-02-00061-a) and European Science Foundation  within activity
``New Trends and Applications of the Casimir Effect''. A.I.V. also thanks the Condensed Matter group of ICTP  for hospitality during the time of working on this article.


\vskip 0.5cm






\begin{thebibliography}{999}

\bibitem{Casimir1948} H.B.G.Casimir, Proc. K. Ned. Akad. Wet., \textbf{51}, 793 (1948).

\bibitem{Lifshitz1955}  E.M.Lifshitz, Zh. Eksp. Teor. Fiz., \textbf{29}, (1955) [Sov. Phys. JETP \textbf{2},73
(1956)]

\bibitem{Rytov1953}   S. M. Rytov,  \textit{Theory of Electrical
Fluctuation and Thermal Radiation} (Academy of Science of USSR
Publishing, Moscow, 1953)

\bibitem{Rytov1967}   M. L. Levin and   S. M. Rytov, \textit{Theory of eqilibrium thermal
fluctuations in electrodynamics} (Science Publishing, Moscow,
1967)

\bibitem{Rytov1989}   S. M. Rytov,  Y. A. Kravtsov, and   V. I. Tatarskii,  \textit{
Principles of Statistical Radiophyics}(Springer, New York.1989),
Vol.3



\bibitem{Dzyaloshinskii1961} I.E.Dzyaloshinskii, E.M.Lifshitz and L.P.Pitaevskii, Adv.Phys. \textbf{10}, 165 (1961)



\bibitem{Lamoreaux2007} S.K.Lamoreaux,   \textit{Phys.Today} \textbf{60}, 40
(Februrary,2007)

 \bibitem{Milloni1994} P.W.Milloni,  \textit{The Quantum Vacuum: An Introduction to Quantum Electrodynamics} (Academic,1993)



\bibitem{NaturePhysics2011} A.O.Sushkov, W.J.Kim, D.A.R.Dalvit and S.K.Lamoreaux,
Nature Phys. \textbf{7}, 230 (2011).

\bibitem{Serry1998} F.M.Serry,   D.Walliser and G.J.Maclay,  J. Appl. Phys. \textbf{84}, 2501 (1998).


\bibitem{Buks2001} E.Buks and M.L.Roukes, Phys.Rev.B \textbf{63}, 033402
(2001).

\bibitem{Klimchinskaya2009} G.L.Klimchitskaya,   U.Mohideen and V.M.Mostepanenko,  Rev. Mod. Phys. \textbf{81}, 1827 (2007).

\bibitem{Munday2009} J.N.Munday, F.Capasso and V.A.Parsegian, Nature (London) \textbf{457}, 170 (2007).

\bibitem{Sushkov2011} A.O.Sushkov,  W.J.Kim,   D.A.R.Dalvit and S.K.Lamoreaux,  Phys. Rev. Lett. \textbf{107}, 171101 (2011).

\bibitem{Bao2010} Y.Bao,  R.Gu\'{e}rout, J.Lussange, A.Lambrecht, R.A. Cirelli,  F.Klemens,  W.M.Mansfield,  C.S.Pai and H.B.Chan,  Phys. Rev. Lett. \textbf{105}, 250402 (2010).


\bibitem{Johnson2011} A.W.Rodriguez,  W.J.Kim,  F.Capasso and S.G.Johnson,  Nature Photon. \textbf{5}, 211 (2011).

\bibitem{Zhao2009} R.Zhao,  J.Zhou,  Th.Koschny,  E.N.Economou and C.M.Soukoulis,  Phys. Rev. Lett. \textbf{103}, 103602 (2009).

\bibitem{Bordag2006} M.Bordag,  B.Geyer,  G.L.Klimchitskaya and V.M.Mostepanenko,  Phys.Rev.B \textbf{74}, 205431 (2006).

\bibitem{Bordag2009} M.Bordag,  I.V.Fialkovsky,  D.M.Gitman and D.V.Vassilevich,  Phys.Rev.B \textbf{80}, 245406 (2009).



\bibitem{Drosdoff2011}  D.Drosdoff and L.M.Woods, Phys.Rev.A \textbf{84}, 062501 (2011).

\bibitem{Drosdoff2010}  D.Drosdoff and L.M.Woods, Phys.Rev.B \textbf{82}, 155459 (2010).

\bibitem{Fialkovsky2011}   I.V.Fialkovsky,  V.N.Marachevsky and D.V.Vassilevich, Phys.Rev.B \textbf{84}, 035446 (2011).

\bibitem{Sernelius2011}   B.E.Sernelius,   \textit{EPL} \textbf{95}, 57003 (2011).

\bibitem{Svetovoy2011}  V.Svetovoy,  Z.Moktadir,  M.Elwenspoek and H.Mizuta,   EPL \textbf{96}, 14006 (2011).




\bibitem{Sarabadani2011}  J.Sarabadani,  A.Naji,  R.Asgari and R.Podgornik, Phys.Rev.B \textbf{84}, 155407 (2011).

\bibitem{Gomez2009}  G.G\'{o}mez-Santos,  Phys.Rev.B \textbf{80}, 245424 (2009).


\bibitem{Geim2004} K.S. Novoselov,  A.K. Geim,  S.V. Morosov,  D. Jiang,  Y. Zhang,  S.V. Dubonos,
I.V. Grigorieva,  and A.A. Firsov,  Science \textbf{306}, 666
(2004).

\bibitem{Geim2005} K.S. Novoselov,    A.K. Geim, S.V. Morosov,  M.I. Katsnelson,  I.V. Grigorieva, S.V. Dubonos,
 and A.A. Firsov,    Nature (London) \textbf{438}, 197 (2005).

\bibitem{Geim2007} A.K. Geim and  K.S. Novoselov,  Nat. Mater. \textbf{6}, 183
(2007).

\bibitem{Geim2009}   Geim A K  \textit{Science} \textbf{324} 1530 (2009)





\bibitem{Pendry1997} J.B. Pendry, J. Phys.C,  \textbf{9}, 10301 (1997).

\bibitem{Shapiro2010} B. Shapiro, Phys. Rev. B  \textbf{ 82}, 075205 (2010)

\bibitem{Volokitin2008b} A.I.Volokitin and  B.N.J. Persson, Phys. Rev. B  \textbf{ 78}, 155437 (2008); \textit{ibid}. \textbf{ 81}, 239901(E) (2010).



\bibitem{Volokitin2001b} A.I.Volokitin and  B.N.J. Persson, J.Phys.:
Condens. Matter  \textbf{13}, 859 (2001).

\bibitem{Mermin1970} N.D.Mermin, Phys. Rev. B  \textbf{ 1}, 2362 (1970).


\bibitem{Wunsch2006} B. Wunscvh,  T. Stauber,   F. Sols, and F. Guinea, New J.Phys.  \textbf{8},318 (2006).


\bibitem{Hwang2007}  E.H. Hwang, S.Das Sarma, Phys. Rev. B \textbf{75}, 205418 (2007).

\bibitem{Chen2008}  J.H. Chen,  C. Jang,  S. Xiao,  M. Ishigami and
M.S.Fuhrer,  Nat. Nanotechnol. \textbf{3},
206 (2008)


\bibitem{Volokitin1999}   A.I.Volokitin and  B.N.J. Persson, J.Phys.:
Condens. Matter  \textbf{11}, 345 (1999); Phys.Low-Dim.Struct.
\textbf{7/8}, 17 (1998).





\bibitem{VolokitinRMP2007}  A.I.Volokitin and  B.N.J. Persson, Rev. Mod. Phys. \textbf{79}, 1291
(2007).

\bibitem{VolokitinPRL2011} A.I. Volokitin   and  B.N.J. Persson, Phys. Rev. Lett.\textbf{106}, 094502 (2011).




\bibitem{Pogrebinskii1977}   M.B. Pogrebinskii,  Fiz.Tekh.Poluprov.  \textbf{11}, 637 (1977)
 [Sov.Phys. Semicond. \textbf{11}, 372 (1977)].

\bibitem{Price1983}   P. J. Price,  Physica B+C \textbf{117},750 (1983).


\bibitem{Gramila1991}   T.J. Gramila,   J.P. Eisenstein,   A.H. MacDonald,
L.N. Pfeiffer, and K. W. West,  Phys. Rev. Lett. \textbf{66}, 1216
(1991).

\bibitem{Sivan1992}   U. Sivan,   P.M. Solomon, and   H. Shtrikman,  Phys. Rev.
Lett.  \textbf{68, }1196 (1992).





\bibitem{Kim2011} S.Kim, I.Jo, J.Nah, Z.Yao, S.K.Banerjee  and E.Tutuc, Phys. Rev. B \textbf{83} 161401 (2011).

\bibitem{Geim2012} R.V.Gorbachev, A.K.Geim, M.I.Katsnelson, K.S.Novoselov, T.Tudorovskyiy, T.V.Grigorieva, A.H.MacDonald, K.Watanabe,
T.Taniguchi  and L.P.Ponamarenko  Nature Phys. \textbf{8} 896 (2012).

\bibitem{Tse2007} W.K.Tse, BenYu-Kuang.Hu and S.DasSarma,  Phys. Rev.B \textbf{76} 081401 (2007).




\bibitem{Katsnelson2011} M.I.Katsnelson, \textit{Phys. Rev.B} \textbf{84} 041407(R) (2011).

\bibitem{Peres2011} N.M.R.Peres, J.M.R.Lopes des Santos and A.H.Castro Neto, Europhys. Lett \textbf{95} 18001 (2011).

\bibitem{Hwang2011} E.H.Hwang, R.Sensarma and S.DasSarma, Phys. Rev. B \textbf{84} 245441 (2011).

\bibitem{Narozhny2012} B.N.Narozhny, M.Titov, I.V.Gornyi  and P.M.Ostrovsky, Phys. Rev. B \textbf{85} 195421 (2012).

\bibitem{Katsnelson2012} M.Carrega, T.Tudorovskiy, A.Principi, M.I.Katsnelson  and M.Polini M 2012  New J. Phys. \textbf{14} 063033 (2012).

\bibitem{Amorin2012} B.Amorin  and N.M.R.Peres,  J. Phys.:Condens. Matter. \textbf{24} 335602 (2012).

\bibitem{Philbin2009} T.G. Philbin and U.  Leonhardt,  New J.
Phys. \textbf{11},033035 (2009).

\bibitem{Pendry2010a} J.B. Pendry,  New J.
Phys. \textbf{12}, 033028 (2010).

\bibitem{Leonhardt2010}  U.  Leonhardt,  New J.
Phys. \textbf{12},068001 (2010).

\bibitem{Pendry2010b} J.B. Pendry,  New J.
Phys. \textbf{12}, 068002 (2010).


\bibitem{Volokitin2011}  A.I.Volokitin and  B.N.J. Persson, New J.
Phys. \textbf{13}, 068001 (2011).

\bibitem{Philbin2011}  T.G.Philbin,
and U.Leonhardt, New J.
Phys. \textbf{13}, 068002 (2011).















\end{thebibliography}
\end{document}